\documentstyle[preprint,prl,aps,psfig]{revtex}
\tightenlines

\begin{document}

\draft

\title{Signals for a Transition from Surface to Bulk Emission in Thermal Multifragmentation}
\author{L. Beaulieu,$^1$ T. Lefort,$^1$ K. Kwiatkowski,$^{1}$\footnote[1]{Present address:  
Los Alamos National Laboratory, Los Alamos, NM 87545} R.T. de Souza,$^1$ W.-c. Hsi,$^1$
L. Pienkowski,$^2$ B. Back,$^3$ D.S. Bracken,$^{1,+}$ H. Breuer,$^{4}$ E. Cornell,$^1$
\footnote[3]{Present address:  Lawrence Berkeley Laboratory, Berkeley, CA 94720}
F. Gimeno-Nogues,$^5$ D.S. Ginger,$^1$\footnote[4]{Cambridge University, Cambridge, U.K.}
S. Gushue,$^{6}$ R.G. Korteling,$^7$ R. Laforest,$^5$\footnote[5]{Barnes Hospital, 
Washington University, St. Louis, MO  63130} E. Martin,$^5$ K.B. Morley,$^2$ 
E. Ramakrishnan,$^5$ L.P. Remsberg,$^{6}$ D. Rowland,$^5$ A. Ruangma,$^5$  
V.E. Viola,$^1$ G. Wang,$^1$\footnote[2]{Present address:  Epsilon, Inc., Dallas, TX 75240} 
E. Winchester$^5$ and S.J. Yennello$^5$}

\address{
$^1$Department of Chemistry and IUCF, Indiana University,
  Bloomington, IN 47405 \\
$^2$Heavy Ion Laboratory, Warsaw University, Warsaw Poland \\
$^3$Physics Division, Argonne National Laboratory, Argonne IL 60439 \\
$^4$Department of Physics, University of Maryland, College Park, MD
20740 \\
$^5$Department of Chemistry \& Cyclotron Laboratory, Texas A\&M
University, College Station, TX 77843 \\
$^6$Chemistry Department, Brookhaven National Laboratory, Upton, NY
11973 \\
$^7$Department of Chemistry, Simon Fraser University, Burnaby,
BC. Canada V5A IS6
}

\date{\today}
\maketitle

\begin{abstract}
Excitation-energy-gated two-fragment correlation functions
have been studied between 2 to 9A MeV of excitation energy for
equilibrium-like sources formed in $\pi^-$ and p + $^{197}$Au 
reactions at beam momenta of 8,9.2 and 10.2 GeV/c. Comparison of
the data to an N-body Coulomb-trajectory code shows a decrease 
of one order of magnitude in the  fragment emission time in the 
excitation energy interval 2-5A MeV, followed by a nearly constant 
breakup time at higher excitation energy. The observed decrease 
in emission time is shown to be strongly correlated with the 
increase of the fragment emission probability, and the onset 
of thermally-induced radial expansion. This result is interpreted 
as evidence consistent with a transition from surface-dominated 
to bulk emission expected for spinodal decomposition.
\end{abstract}

\pacs{25.70.Pq,21.65+f,25.40-h,25.80.Hp}

%paper starts here

The nuclear liquid-gas phase transition is often linked to the 
experimental observation of multiple intermediate mass fragments 
(IMF=  3 $\leq$ Z $\stackrel{=}{\sim}$ 20), or multifragmentation
\cite{mor93}. This interpretation was preceded by 
the discovery of a power-law behavior in the mass (and charge) 
distributions of IMFs in high energy p+Xe
reactions~\cite{Fin82}. In recent years, better experimental 
technologies have made possible the detection 
and nearly complete characterization of multifragmentation reactions 
on an event-by-event basis. Two {\em stimulating} results have come 
from such exclusive experiments: the measurement of a latent heat 
for nuclear matter by Pochodzalla et al.~\cite{Poc95},
reminiscent of the liquid-gas phase transition in water,and 
the extraction of critical exponents for nuclear matter, in agreement 
with those of liquid-gas values, by the EOS collaboration~\cite{Gil94}.
While debate continues related to these findings~\cite{EOS,Kri98,INDRA,TAMU}, 
the important point remains that the current state of the data is 
suggestive of a liquid-gas phase transition, either first or second order. More 
recently, it has been argued that due to finite-size effects, both 
first- and second-order phase transition characteristics can be 
observed simultaneously~\cite{Cho99}.

An aspect of this phase transition that has been less systematically
examined is the cluster-emission time scale involved in the transition 
of the nuclear liquid to a liquid-vapor coexistence phase.  
For cluster emission from the liquid phase, fragment formation occurs 
at the surface of the excited source via a binary splitting, much like 
fission. This process gives rise to long emission times of the order 
of 10$^{-20}$ - 10$^{-21}$ sec, necessary for shape deformation.
In sharp contrast, when the spinodal boundary of the phase diagram 
is crossed, the system falls apart on a very short time scale 
($\sim$ 10$^{-22}$ - 10$^{-23}$sec); i.e. bulk emission. 

Although the transition from long to short emission times has previously 
been reported for central heavy-ion collisions, no 
systematic analysis of source lifetime as a function of excitation
energy per nucleon, $E^*/A$, has yet been performed for a single, 
well-defined equilibrated system. Previous studies have estimated the 
excitation energy scale by either fixing the target-projectile
combination and varying the bombarding energy~\cite{bau93}, or
combining results from several different experiments, in which $E^*/A$ 
was evaluated~\cite{Dur98}. In both cases, there are problems in heavy-ion 
reactions associated with entrance-channel dynamics and with source 
selection, which lead to a single thermalized source for less than 
few percent of the total reaction cross-section\cite{Bea96,Mar98,Tok96}.

In this letter we present the systematic evolution 
of IMF emission times over a wide range of excitation energy for 
well characterized, thermal-like sources formed in hadron-induced 
reactions, as described in~\cite{Lef99,Bea99}. This evolution is
then correlated to the n-fold emission probability and to the onset of 
thermally-induced radial expansion. The convergence of the results 
suggests a transition from surface emission to bulk emission around 
5A MeV of excitation energy, in good agreement with the theoretical 
prediction of Bondorf et al.~\cite{Bon85}. 

Emission times were obtained using excitation-energy-gated 
two-IMF correlation functions constructed from events
measured with the ISiS 4$\pi$ detector array~\cite{Kris/ISiS}
during experiment E900 at the Brookhaven National Laboratory 
AGS accelerator. Beams of 8.0 GeV/c tagged $\pi^-$and 8.2, 9.2 GeV/c 
$\pi^-$ and 10.2 GeV/c protons (untagged) were incident on a 1.8 
mg/cm$^2$ $^{197}$Au target.
Fragments with charge Z$\leq$16 were identified with a set of
162 gas-ion chamber/Si/CsI triple telescopes with energy acceptance
1.0$\leq$E/A$\leq$92 MeV. Geometric acceptance was 74\% of 4$\pi$. 
Further experimental details can be found in~\cite{Lef99,Hsi97}.
For all these reactions, the reconstructed primary source size,
final residue size, charge distribution, fragment multiplicity
and cross-section were found to be the same when selected as a function of
$E^*/A$~\cite{Bea99,Hsi97}. By combining these data sets, we
have access to 7 million events for which $\sim$7\% contain
two or more IMFs with 4$\leq Z_{imf} \leq$9.

Emission time scales were derived using the intensity-interferometry
technique~\cite{Boa90,tro87,bou89,kim91,lou95,fox94,gla94,Pop98,Wan99}, which 
employs two-fragment reduced-velocity correlation functions, defined as

\begin{equation}
R(v_{red})+1=C\frac{N_{corr}(v_{red})}{N_{uncorr}(v_{red})},
\end{equation}

\noindent where $v_{red}$ is the reduced relative velocity
between the two fragments:

\begin{equation}
v_{red}=\frac{\mid \vec{v_1}-\vec{v_2} \mid}{\sqrt{Z_1+Z_2}}.
\end{equation}

\noindent Here $v_i$ and $Z_i$ are the laboratory velocity
and charge of the fragments, respectively. The denominator permits
comparison of different $Z$ values~\cite{kim91}.
$N_{corr}(v_{red})$ is the measured coincidence yield while
$N_{uncorr}(v_{red})$ is the uncorrelated yield calculated
using the event-mixing technique~\cite{kim91}. For each $E^*/A$ 
bin, both the numerator and demominator were normalized to their 
integral and then the ratio was constructed. At large $v_{red}$, constant values 
of $R(v_{red})+1 \approx$ 1 are found, largely due to the use of the 
full 4$\pi$ range of the array, which enables high statistical 
samples of large-angle correlations.

The experimental excitation-energy-gated IMF-IMF correlation 
functions (4$\leq Z_1,Z_2 \leq$9) are shown in Fig. 1 for source 
excitation energies $E^*/A$= 2.25 $\pm$ 0.25, 3.0 $\pm$ 0.5, 5.0$\pm$ 1.0 
and 8.0 $\pm$ 1.0 MeV. Yield suppression at low $v_{red}$, due
to the Coulomb interaction between IMFs, increases with $E^*/A$. The increase
of the Coulomb hole (yield suppression) is large between $E^*/A$ = 2.25 MeV 
and 5 MeV of excitation energy, followed by a nearly constant yield 
suppression for higher excitation energies. This effect 
is in agreement with previous analyses that have selected the most 
central collisions of a given reaction and varied the bombarding 
energy~\cite{bau93,lou95}. However, as discussed in the introduction, 
the formation of a single source in heavy-ion collisions is limited to 
a fraction of the cross-section, which makes the task of separating 
the actual thermalized source from contamination difficult. In hadron-induced 
reactions, there is only one source of thermalized fragments, it has 
little collective compressional or rotational character, and the data 
cover a wide range of excitation energy.

The evaluation of the emission time scale at various excitation
energies is  done by comparison of the data with the $N$-body Coulomb
trajectory code of Glasmacher {\it et al}~\cite{gla94,Pop98}. 
Because the starting source size, as well as the final residue size, are 
known, the simulation is  simplified; i.e. no ``empirical'' adjustments 
of these quantities is possible, only the volume or separation distance 
between the residue and the fragments. The Coulomb-trajectory simulation 
employed in this time-scale analysis assumes sequential emission from the 
surface of a spherical source; thus there are no complications due to 
initial-state momentum correlations~\cite{Pop98}.

Two-dimensional maps of measured source charge vs. residue charge and IMF 
kinetic energy in source frame vs. charge are used to define the initial conditions.
The filtered output of the simulation is required to reproduce both the 
small- and large-angle correlation data, as well as fragment kinetic-energy 
spectra and charge distributions. As in~\cite{Wan99}, average 
Coulomb-barriers are subtracted at the input for each $Z$, effectively
sampling the exponential (thermal) part of the spectra. The Coulomb energy 
is calculated and added back in the simulation according to 

\begin{equation}
V_{Coul} ={ {1.44  Z_{IMF}\:  Z_{res} } \over
    {r_o(A^{1/3}_{IMF}\:+\: A^{1/3}_{res})+d}} ,
\end{equation}

\noindent where $r_o$ = 1.22 fm and $d$ is an adjustable parameter that
defines the source dimensions. At low excitation, a distance of 6 fm is 
required to reproduce the energy spectra down to 2A MeV, corresponding to 
IMF-residue axial separation distances consistent with fission systematics 
using $r_0$=2 fm and $d$=0 fm in Eq. 3. At high excitation, values of $d$ as 
low as 2 fm will reproduce the energy spectra down to 1A MeV. For the purpose 
of comparison of the data with the calculations, correlation functions for 
IMF kinetic energy between 2A MeV and 10A MeV are used.  Equivalently, assuming 
radial separation instead of axial (high excitation scenario), the 
above range for $d$ (2-6 fm) would correspond to a freeze-out volume of 
4$V_0$-6$V_0$ (2.5$V_0$-4$V_0$) using $r_0$=1.22 fm (1.44 fm) as a normal 
density reference.  Finally, the emission time $t$ is assigned via an 
exponential probability distribution, e$^{-t/\tau}$, where $\tau$ is the decay lifetime.

In Fig.~\ref{fig2} we show fits to the correlation functions for three
bins in $E^*/A$ for a range of d and $\tau$ that yield minimum chi-squared 
values.  Between $E^*/A$ = 2 - 2.5 MeV and 4.5-5.5 MeV, the emission time 
decreases nearly an order of magnitude from $\sim$500 fm/c to 30-75 fm/c.  
Above $E^*/A$  $\sim$ 5 MeV, the emission time becomes very short 
(20 to 50 fm/c) and nearly independent of excitation energy, 
consistent with a near-instantaneous breakup, explosion-like phenomenon.  
As has been pointed out previously, lifetimes less than $\tau \sim$30 
fm/c become comparable to thermodynamic fluctuations in the system, 
as discussed in ref.~\cite{Wan99}. Moreover, correlation functions
for a similar reaction have been shown to be reproduced by a simultaneous
multifragmentation model at high excitation~\cite{Wan99}, again consistent
with our extracted time scale.
 
As in past studies, Fig.~\ref{fig2} provides evidence for the space-time
ambiguity of the correlation function~\cite{fox94}. However, the difference
between the various decay lifetimes at high $E^*/A$ is rather 
small and agrees within the error bars. Therefore, the observed
saturation in the space-time extent of the source reflects to
a large degree a saturation of the decay lifetime as well.

The lower frame of Fig.~\ref{fig3} presents the best-fit decay times 
for hadron-induced thermal multifragmentation of $^{197}$Au nuclei, 
with two extreme solutions shown above 4A MeV. The decay lifetimes evolve 
systematically as a function of $E^*/A$, extending from the evaporative 
regime at low $E^*/A$ to that for multifragmentation at high $E^*/A$. 
The shaded band shows the range 
of space-time values for which a consistent fit to all of the observables 
is achieved. The solid points represent independent measurements for 
heavy-ion-induced reactions for which the source $E^*/A$ has been explicitly 
evaluated~\cite{Dur98}. At high excitation energies the ISiS results are 
similar to the heavy-ion data. However, as $E^*/A$ decreases, longer 
lifetimes are derived from the heavy-ion data. The differences 
can possibly be explained by a better source selection in hadron-induced reaction 
with minimal influence from various entrance-channel collective behavior, 
all of which contribute to lowering the reconstructed $E^*/A$~\cite{Bea99}.

The solid line in the lower frame of Fig. 3 shows the best fit to an 
exponential function ($e^{\alpha/\sqrt{E^*/A}}$) for the ISiS data; 
the dashed line describes a similar fit using the heavy-ion results 
at low excitations. These two lines should be seen as defining upper and 
lower limits of the changes in emission time with $E^*/A$ based on all 
available data. 

In ref.~\cite{Lef99b} the thermally-driven radial expansion, $\epsilon_r$,
has been derived for the ISiS data as a function of $E^*/A$, based on
an analysis of the fragment kinetic energy spectra. The center frame in 
Fig.~\ref{fig3} shows that onset of thermal expansion energy for the ISiS 
data, which occurs around $E^*/A$=3.5 MeV (4.75 MeV), assuming a freeze-out 
density of $\rho/\rho_0$=1/3 (1/2). The measurements of a distinct signal for 
$\epsilon_r$ is interpreted as evidence that the system has reached a reduced 
density~\cite{Lef99b}, an essential condition for bulk emission.

In the upper panel of Fig. 3 the decay lifetime systematics are compared 
with the probability for a given observed IMF multiplicity, 
N$_{IMF}$, as a function of $E^*/A$, uncorrected for ISiS efficiency.
The dotted vertical lines that run across all three frames -- $E^*A$ 
4.25$\pm$0.50 MeV -- represent an apparent transition region in which 
multiple IMF emission becomes the dominant process, the onset of 
thermally-induced radial expansion appears, and decay times become 
comparable to the thermodynamic fluctuation time ($\tau$ $<$ 30 fm/c). 
Furthermore for the same system under study, this region in $E^*/A$ yields 
the most diverse fragment size distribution (minimum power-law 
exponent)~\cite{Bea99}. We interpret 
the simultaneous change of the various observables in this region 
of $E^*/A$ as evidence supporting a transition from a surface emission 
mechanism (long timescales) to bulk disintegration (short timescales) 
in hot nuclear matter. 

In summary, IMF-IMF correlation functions for hadron-induced thermal 
multifragmentation events have been studied in the excitation energy
regime $E^*/A$ = 2-9 MeV.  The Coulomb suppression of the correlation
function at low reduced velocities shows a systematic evolution with 
increasing heat content. Long times (minimal suppression) are 
associated with low excitation energies and short times with larger 
values (large suppression).  Between $E^*/A$ = 2 - 5 MeV, this evolution is quite
strong; above 5 MeV/A the correlation functions show little change.
Decay lifetimes that decrease from $\tau \sim$ 500 fm/c at $E^*/A$ = 2
MeV to $\tau \sim$ 20-50 fm/c for $E^*/A$ $\geq$ 5 MeV are derived from
fits to the correlation functions with an N-body Coulomb-trajectory 
simulation that also reproduces the IMF kinetic-energy spectra, charge
distributions and large-angle correlations.  Placing this timescale
evolution with $E^*/A$ in context with similar behavior for multiple
fragment production probabilities and the onset of thermally-induced
radial expansion, we conclude that the data provide evidence consistent 
with a transition from surface to bulk emission in hot nuclear matter 
near $E^*/A$ = 5 MeV.

The authors wish to thank Thomas Glasmacher for providing his
N-body Coulomb-trajectory code. This work was supported by the U.S. 
Department of Energy and National Science Foundation, the National 
Sciences and Engineering Research Council of Canada, Grant No. P03B 
048 15 of the Polish State Committee for Scientific Research, Indiana 
University Office of Research and the University Graduate School, Simon 
Fraser University and the Robert A. Welch Foundation.

% now the references. delete or change fake bibitem. delete next three
%   lines and directly read in your .bbl file if you use bibtex.

% \begin{figure}
% \caption{}
% \label{}
% \end{figure}

\begin{figure}
\caption{Reduced velocity correlation functions generated for four
  different excitation energy per nucleon bins. IMF kinetic energy 
  acceptance is in source frame is E$_{IMF}$/A = 1 - 10 MeV.}
\label{fig1}
\end{figure}

\begin{figure}
\caption{Correlation functions for Z = 4-9 IMFs as a function of
  reduced velocity (open circles).  IMF kinetic energy acceptance in the source 
  frame is E$_{IMF}$/A = 2 - 10 MeV.  Data gated on source E*/A = 2.0 - 2.5 MeV
  (top), 4.5 -5.5 MeV (center)and 8.5 - 8.5 MeV (bottom). Solid and
dashed lines are results of a Coulomb trajectory calculation for fit
parameters indicated on figure.}
\label{fig2}
\end{figure}

\begin{figure}
\caption{Dependence on $E^*/A$ for source lifetime (bottom), thermally-driven
expansion energy $\epsilon_r$~\protect\cite{Lef99b} (center) and probability of 
observing a given IMF multiplicity (top). In the bottom panel, the shaded area 
indicates the range of possible solutions (space-time) consistent with 
IMF observables. Solid line is an exponential fit to the ISiS results; 
dashed line a similar fit using heavy-ion data~\protect\cite{Dur98}.}
\label{fig3}
\end{figure}

\newpage

\begin{figure}
\centerline{\psfig{file=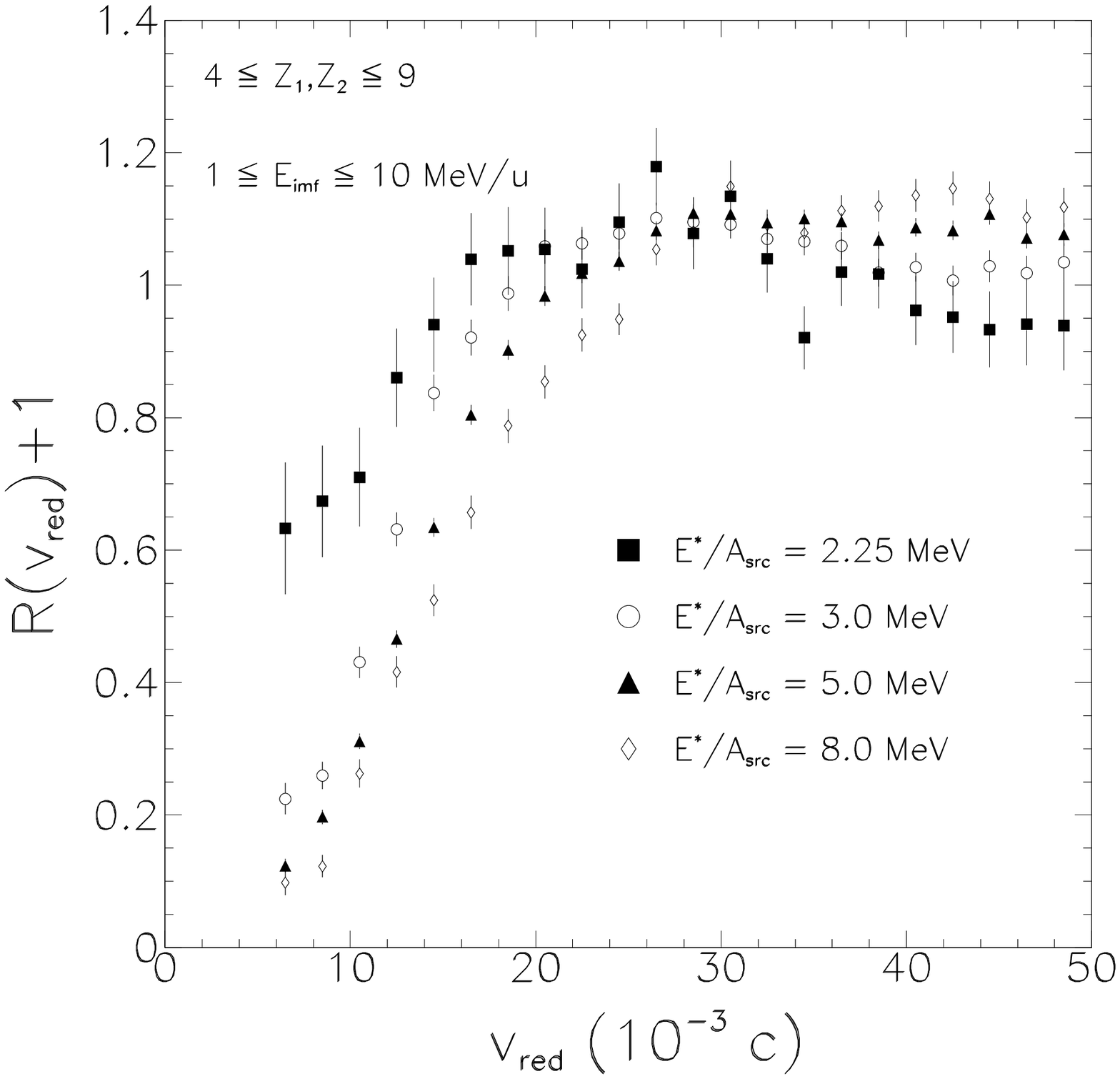,width=7.5in}}
\vspace{1in}
\center{Fig. 1: L. Beaulieu {\em et al.}}
\end{figure}

\newpage

\begin{figure}
\centerline{\psfig{file=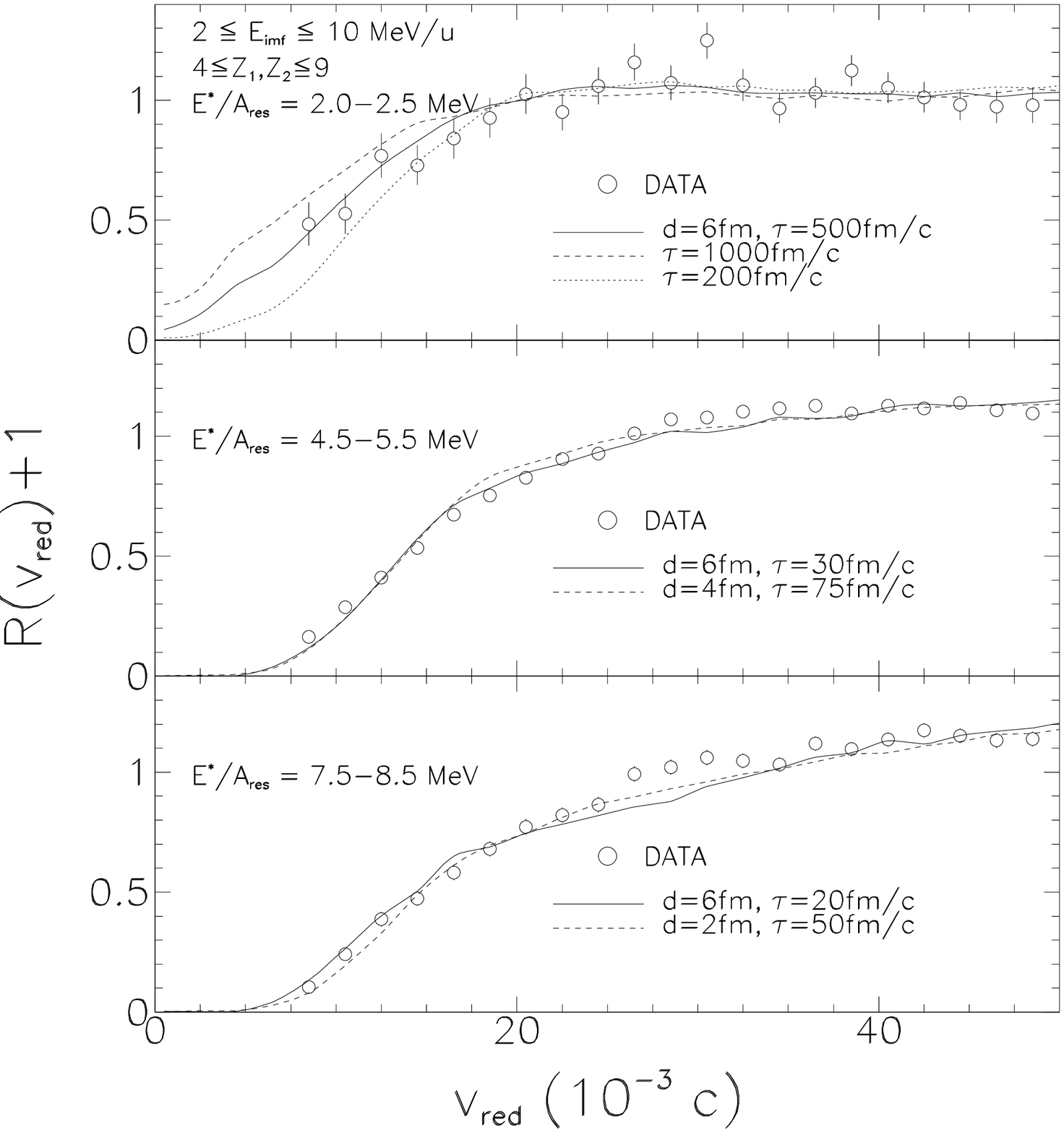,width=6.0in}}
\vspace{0.2in}
\center{Fig. 2.: L. Beaulieu {\em et al.}}
\end{figure}

\newpage

\begin{figure}
\centerline{\psfig{file=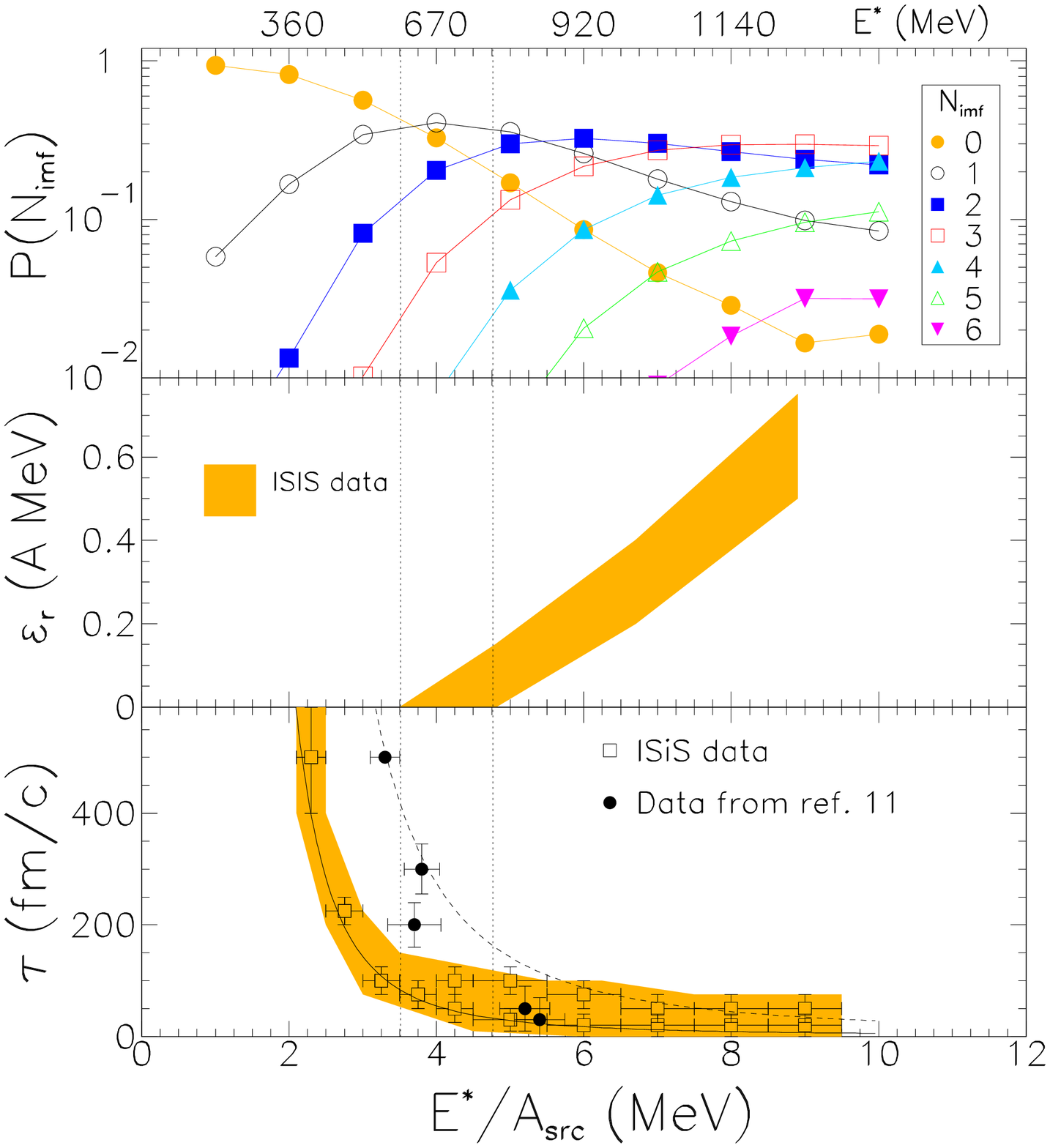,width=6.0in}}
\vspace{0.2in}
\center{Fig. 3: L. Beaulieu {\em et al.}}
\end{figure}

\end{document}